\RequirePackage{ifpdf}
\documentclass{PoS_oldRef}

\title{Multiple supermassive black hole systems: \newline SKA's future leading role}

\ShortTitle{Multiple supermassive black hole systems}

\author{\speaker{Roger Deane}$^{1,2}$, Zsolt Paragi$^{3}$, Matt Jarvis$^{4,5}$, Mick\"ael Coriat$^{1,2}$, Gianni Bernardi$^{2,6,7}$, Sandor Frey$^{8}$, Ian Heywood$^{9,6}$, Hans-Rainer Kl\"ockner$^{10}$ 

\\ 

$^1$ University of Cape Town, 
$^2$ Square Kilometre Array South Africa, 
$^3$ Joint Institute for VLBI in Europe, 
$^4$ University of Oxford, 
$^5$ University of the Western Cape,
$^6$ Rhodes University, 
$^7$~Harvard-Smithsonian Center for Astrophysics, 
$^8$ F\"OMI Satellite Geodetic Observatory, 
$^9$ CSIRO Astronomy and Space Science, 
$^{10}$ Max-Planck-Institut f\"ur Radioastronomie
 
\\
E-mail: \email{roger.deane [at] ast.uct.ac.za}
}

\abstract{Galaxies and supermassive black holes (SMBHs) are believed to evolve through a process of hierarchical merging and accretion. Through this paradigm, multiple SMBH systems are expected to be relatively common in the Universe. However, to date there are poor observational constraints on multiple SMBHs systems with separations comparable to a SMBH gravitational sphere of influence (<< 1 kpc). In this chapter, we discuss how deep continuum observations with the SKA will make leading contributions towards understanding how multiple black hole systems impact galaxy evolution. In addition, these observations will provide constraints on and an understanding of stochastic gravitational wave background detections in the pulsar timing array sensitivity band (nHz -$\mu$Hz). We also discuss how targets for pointed gravitational wave experiments (that cannot be resolved by VLBI) could potentially be found using the large-scale radio-jet morphology, which can be modulated by the presence of a close-pair binary SMBH system. The combination of direct imaging at high angular resolution; low-surface brightness radio-jet tracers; and pulsar timing arrays will allow the SKA to trace black hole binary evolution from separations of a galaxy virial radius down to the sub-parsec level. This large dynamic range in binary SMBH separation will ensure that the SKA plays a leading role in this observational frontier.  }

\FullConference{
Advancing Astrophysics with the Square Kilometre Array\\
June 8-13, 2014\\
Giardini Naxos, Sicily, Italy}

\newcommand{\skipthis}[1]{}
\newcommand{\mnras}{MNRAS}
\newcommand{\nat}{Nature}

\newcommand{\apj}{ApJ}
\newcommand{\apjl}{ApJL}
\newcommand{\aap}{A\&A}

\newcommand{\araa}{ARA\&A}

\newcommand{\nar}{New Ast. Reviews}

\begin{document}

\section{Introduction}

Binary supermassive black hole (SMBH) systems have long been predicted to be common in the Universe (Begelman, Blandford \& Rees, 1980). Simulations suggest that they have a broad range of astrophysical impacts, including shallowing the inner density profiles of galactic halos as they eject stars via three-body interactions during in-spiral (e.g. Graham 2004, Merritt 2006); and an increase in bulge star formation and black hole accretion through disruption of cold gas angular momentum (e.g. Blecha et al.~2011, van Wassenhove et al.~2012). Furthermore, sub-parsec binary SMBHs are expected to dominate the stochastic gravitational wave background at nHz-$\mu$Hz frequencies (Wyithe \& Loeb 2003, Sesana 2013). Numerical simulations of the large-scale structure formation of the Universe reveal a process of hierarchical galaxy merging (e.g. Springel et al.,~2005). Since every galaxy is expected to host a SMBH (Kormendy \& Richstone 1995), each galaxy merger should include a merger of SMBHs. Despite this forecasted ubiquity and the broad range of predicted binary SMBH impacts, our observations of these systems remain very limited. Recently, there has been a resurgence in effort to find more dual/binary AGN\footnote{We call a SMBH system dual if the separation of its components is larger than their radii of gravitational influence, and binary otherwise (see Merritt 2013).}, perhaps most notably by a hard X-ray census of the local Universe (Koss et al. 2012); a large-scale search through VLBI data archives for double flat-spectrum sources (Burke-Spolaor 2011); and 2D spectroscopic and near-infrared, adaptive-optics-assisted followup imaging of double-peaked narrow emission line AGN selected from SDSS (e.g. Rosario et al. 2010, 2011; McGurk et al. 2011; Fu~et~al.~2011, Comerford et al. 2011). This has resulted in an increase in the number of dual AGN on $\sim$1-100~kpc scales, however, to date there are only four strong candidate sub-kpc binary/dual AGN systems (Komossa et al. 2003, Rodriguez et al. 2006, Fabbiano et al. 2011, Deane et al. 2014). In this chapter we discuss how the SKA will take a leading role in exploring this observational frontier.

\subsection{Observational status}

\begin{figure} 
\centering
\includegraphics[width=0.95\textwidth]{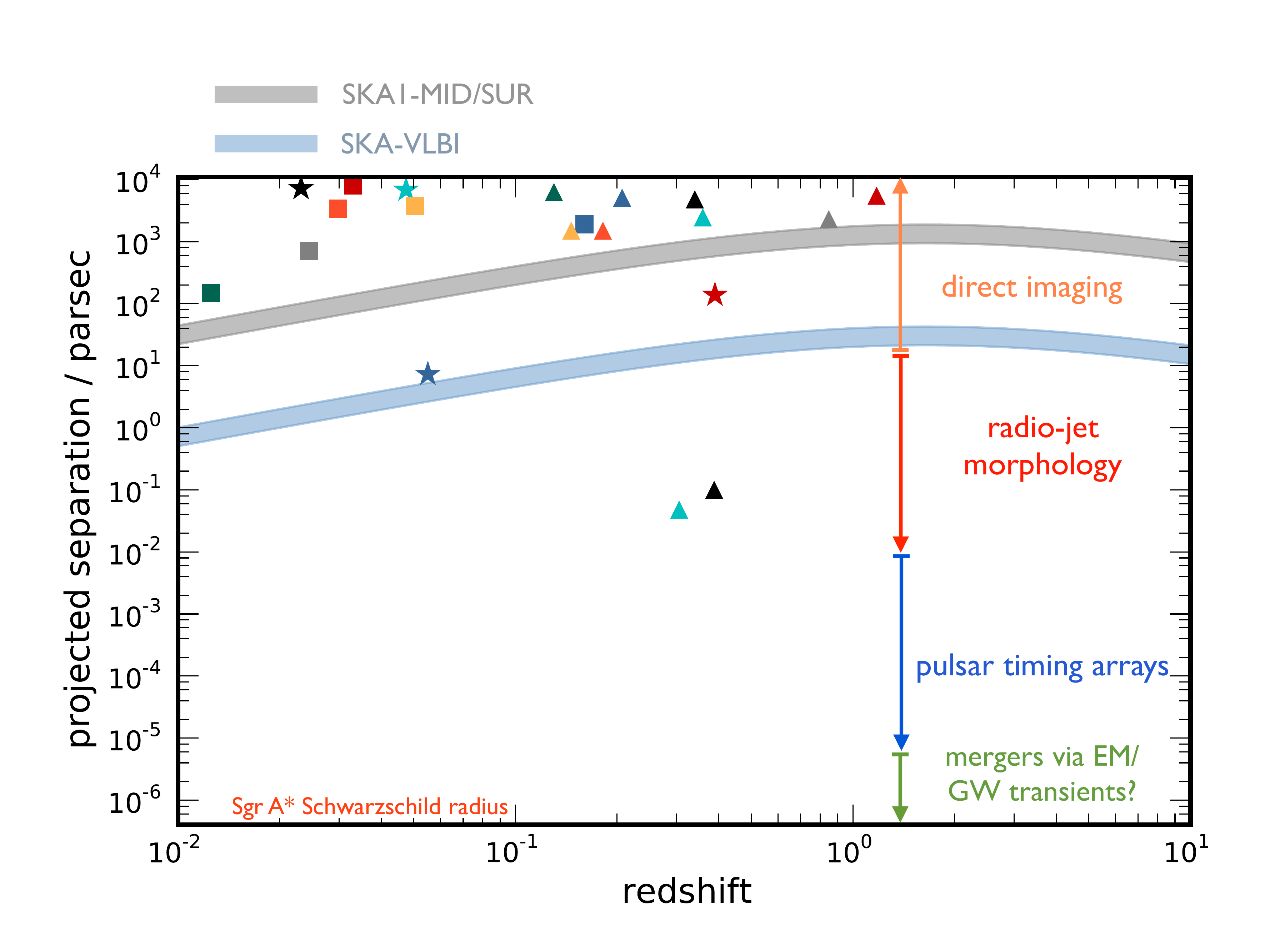} 
\caption{ Sample of the strongest dual/binary AGN candidates as revealed by direct imaging at X-ray (squares), optical/infrared (triangles) and radio (stars) wavelengths. The two sub-pc triangles denote two spatially-unresolved candidates based on their double-peaked broad emission lines and quasi-periodic light curve. The grey and blue curves indicate the spatial resolution (for frequency range 1-2 GHz) that will be possible with SKA1-MID/SUR and SKA-VLBI. The labels on the right provide a rough sketch on how the SKA will explore $\gtrsim$6 decades of binary SMBH orbital separation and are elaborated on later in this contribution. The projected separation lower limit is set to the Schwarzschild radius of Sgr A* ($r_{\rm s} \sim 4 \times 10^{-7}$~pc). Source names can be found in Deane et al.~(2014)}
\label{fig:status}
\end{figure}

In Fig.~\ref{fig:status} we plot what could be considered as the strongest candidate binary/dual SMBH systems. While there may be a degree of subjectivity on which sources should appear in this figure, it illustrates how few candidates there are, despite our expectations to the contrary. Nonetheless, progress has been made in the past decade considering that only three systems in this figure could be found in the literature at the time the first SKA science case was published (Carilli \& Rawlings, 2004). In the next decade, large-scale surveys with the SKA will make significant contributions towards populating the sub-kpc parameter space in particular, driven by superior angular resolution and sensitivity, negligible dust and gas attenuation at GHz frequencies, and the enhanced nuclear accretion that appears to take place in kpc-scale dual and triple AGN (see Sec.~\ref{sec:enhancedjet}). In Fig.~\ref{fig:status} we show the approximate angular resolution at 1-2~GHz of SKA1-MID/SUR and SKA-VLBI (see Paragi et al.~2014, these proceedings), which illustrates its advantage to search for these systems, particularly when considering the wide-area surveys the SKA will perform in this frequency band.

\subsection{Impact on Galaxy Evolution}\label{sec:GE}

The prevalence and evolution of multiple SMBH systems are predicted to have a broad range of astrophysical impacts, which have important galaxy evolution implications. Gravitational perturbations from binary systems are expected to increase both SMBH accretion and bulge star formation rates, and hence impact the well-correlated nuclear black hole to spheroid mass ratio (e.g. McConnell \& Ma 2013). Additional scatter in this relation may result from the potential ejection of recoiling black holes (e.g. Blecha et al.~2011). Hydrodynamical simulations suggest that the separations of $\sim$10$^8$~M$_\odot$ binary SMBHs can decrease from $\sim$100 parsec down to the sub-pc gravitational radiation dominated regime on a timescale of a few Myr if sufficient gas is present (e.g. van Wassenhove et al.~2012). As the orbital separation decreases, the binary SMBH system is predicted to eject of order 1-4 times their combined mass from the galaxy/bulge via three-body interactions (Merritt 2006). This mass is sufficient to flatten the inner density slope, as is measured in nearby galaxies (Merritt \& Milosavljevi{\'c} 2005). Therefore, this appears to be an important consideration in the attempts to reconcile theoretically motivated central matter density cusps in a $\Lambda$CDM cosmology versus the observationally supported cores which have flat central density profiles (e.g. Mashchenko et al. 2006). The anisotropic gravitational wave emission from binary coalescence is predicted to result in large recoil velocities and should enable the measurement of AGN offset from the host galaxy centre (e.g. Madau \& Quataert 2004). However, apart from a number of strong individual candidates (e.g. Civano et al. 2010), a detailed analysis of 14 nearby cored elliptical galaxies suggests this offset (where measured) is not as large as predicted for plausible merger rates (Lena et al. 2014).

\subsection{Linking with pulsar timing array experiments}\label{sec:gw}

While binary SMBHs are expected to dominate the low-frequency stochastic gravitational wave background, very little is known about the properties of low separation binaries themselves, such as the typical binary in-spiral rate, eccentricity evolution and environmental coupling at sub-kpc scales. These are important to constrain as they directly determine the low-frequency gravitational wave spectral normalization and shape. Indeed, stochastic gravitational wave background predictions typically assume that nature solves the so-called `last parsec problem'. This arises from the estimate that binaries take of order a Hubble time to merge via gravitational radiation, following the ejection of most matter within binary orbital separations of $\sim$1~parsec (Merritt \& Milosavljevi{\'c} 2005). Statistics from a large sample of binaries will measure the in-spiral rate and directly address the question of whether binaries `stall' or not. However, more insight on eccentricity evolution will require further successes in the sophisticated simulations that have developed in the recent past (e.g. Mayer et al.~2007, Blecha et al.~2011, Kulkarni et al.~2012, van Wassenhove et al.~2012) as well as detailed ALMA observations. Stellar scattering driven models predict that if typical binary SMBHs have an initial eccentricity of $e_0 \sim 0.7$ at formation, the expected characteristic strain at 1 nHz is suppressed by a factor of $\sim$5, while the effect is minimal at higher frequencies ($\sim$1-10 $\mu$Hz, Sesana 2013, Ravi et al.~2014). Moreover, the presence of significant gas masses appears to increase the eccentricity (although, perhaps not to the extremely high values in the stellar driven case). Such eccentric systems are predicted to emit sharp spikes of gravitational wave radiation as the black holes reach the pericentre of their increasingly eccentric orbit. For eccentricities of $e \sim 0.9-0.95$, the characteristic strain will increase by 2-3 orders of magnitude for a constant semi-major axis. This implies that the highly eccentric inner binaries that result in simulated triple systems in particular (e.g. Hoffman \& Loeb 2007) may result in detectable sources for pointed gravitational wave experiments (for a small fraction of their orbit). While this may be promising for detecting gravitational wave hotspots, it is likely to decrease the amplitude of the stochastic gravitational wave spectrum at nHz frequencies by significant factors. Therefore, the combination of a large sample of SKA-discovered binaries; the continued progress on sophisticated simulations; as well as molecular gas dynamics at high angular resolution with ALMA are important in understanding the characteristic binary in-spiral evolution and resultant nHz gravitational wave spectrum. See Burke-Spolaor (2013) for an excellent review on the multi-messenger astrophysics that will be ushered in by linking pulsar timing array results with electromagnetic probes.

\section{Motivation: why the SKA will lead}

The combination of depth, area and angular resolution of SKA continuum surveys will result in an unprecedented view of the faint radio Universe, and by extension, multiple SMBH systems. However, there are a number of attributes of these hitherto exotic systems that will further enable their detailed study with the SKA.

\subsection{Enhanced radio-jet triggering in dual and triple AGN}\label{sec:enhancedjet}

\begin{figure}
\centering
\includegraphics[width=0.6\textwidth]{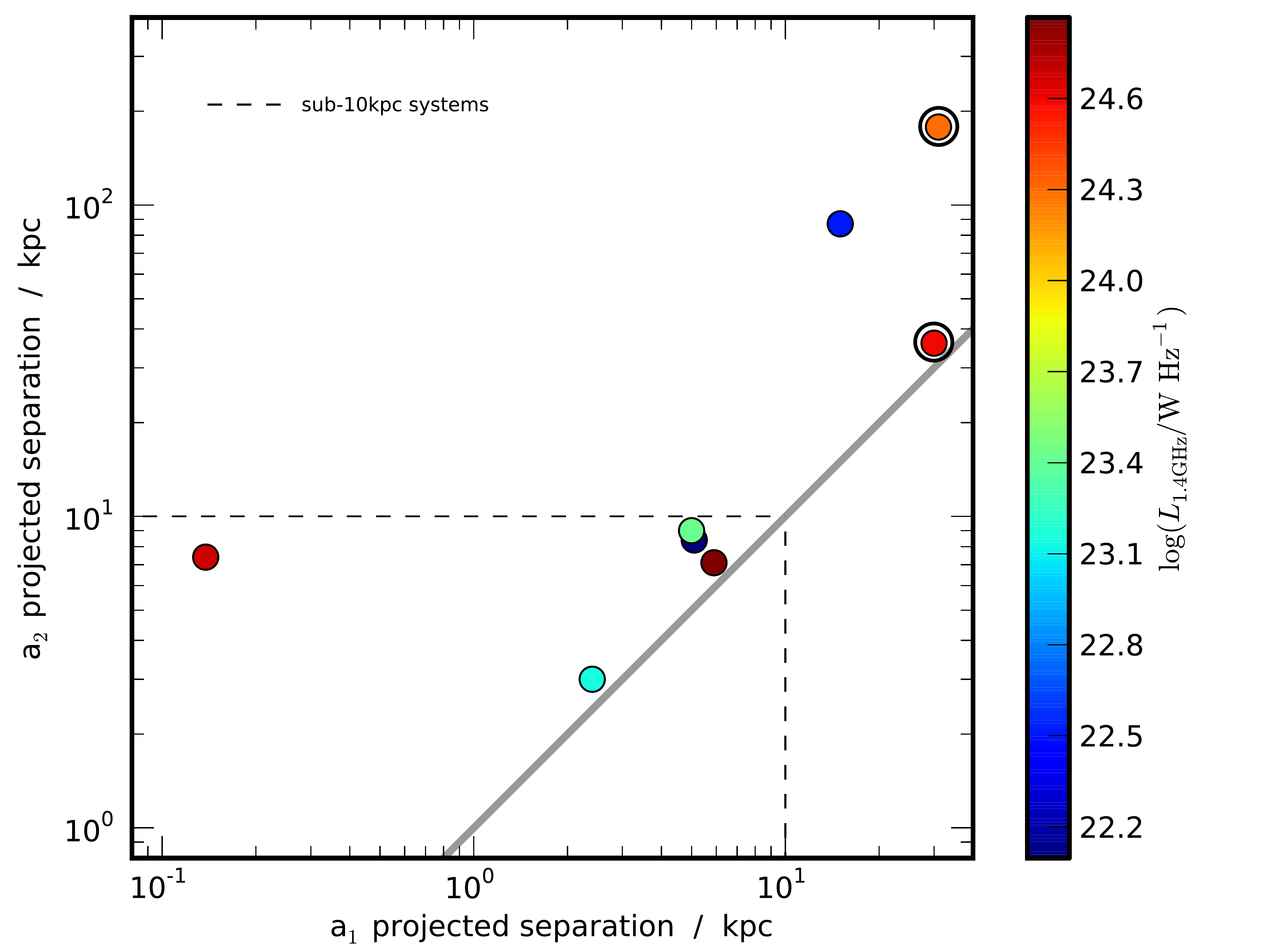} 
\caption{ Projected separations of candidate triple active galactic nuclei. The axes denote the lowest and 2$^{\rm nd}$ lowest separations between AGN. The solid black line indicates a ratio of unity. The dashed lines isolate the sub-10kpc systems, which is roughly the effective stellar radius of an elliptical galaxy. The colours correspond to the total integrated radio luminosity of the AGN plus the host galaxy, since low angular resolution limits a fair AGN radio luminosity comparison. Nonetheless, the plot supports enhanced accretion within triple systems, particularly in the sub-10kpc systems. The two black circled points (which are triple quasar systems) indicate upper radio luminosity limits from archival observations. Figure reproduced from Deane~et~al.~(2014), with references therein. }
\label{fig:triples}
\end{figure}

It has been shown that galaxy-galaxy interactions are 
extremely efficient at triggering both nuclear star formation (e.g. Scudder 
et al. 2012; Patton et al. 2013) and AGN activity (e.g. Ellison et al. 2011), 
and that these trends continue well into the post-merger phase (Ellison et 
al. 2013). These results are consistent with predictions from hydrodynamical simulations (e.g. van Wassenhove et al.~2012, Blecha et al.~2011) and this implies that radio-jet triggering should become more prevalent amongst binary and triple SMBHs (see Fig.~\ref{fig:triples}) . Following this (admittedly simplistic, yet plausible) argument would lead us to expect the efficiency of binary SMBHs to increase with decreasing orbital separation, provided sufficient angular resolution and survey sensitivity. If correct, the sub-arcsecond angular resolution and $ \lesssim \, 1\,\mu$Jy\,beam$^{-1}$ wide-area SKA surveys will undoubtedly enable the efficient discovery of a large population of sub-kpc binary/dual AGN.

\begin{figure}
\begin{center}
\includegraphics[width=0.7\textwidth]{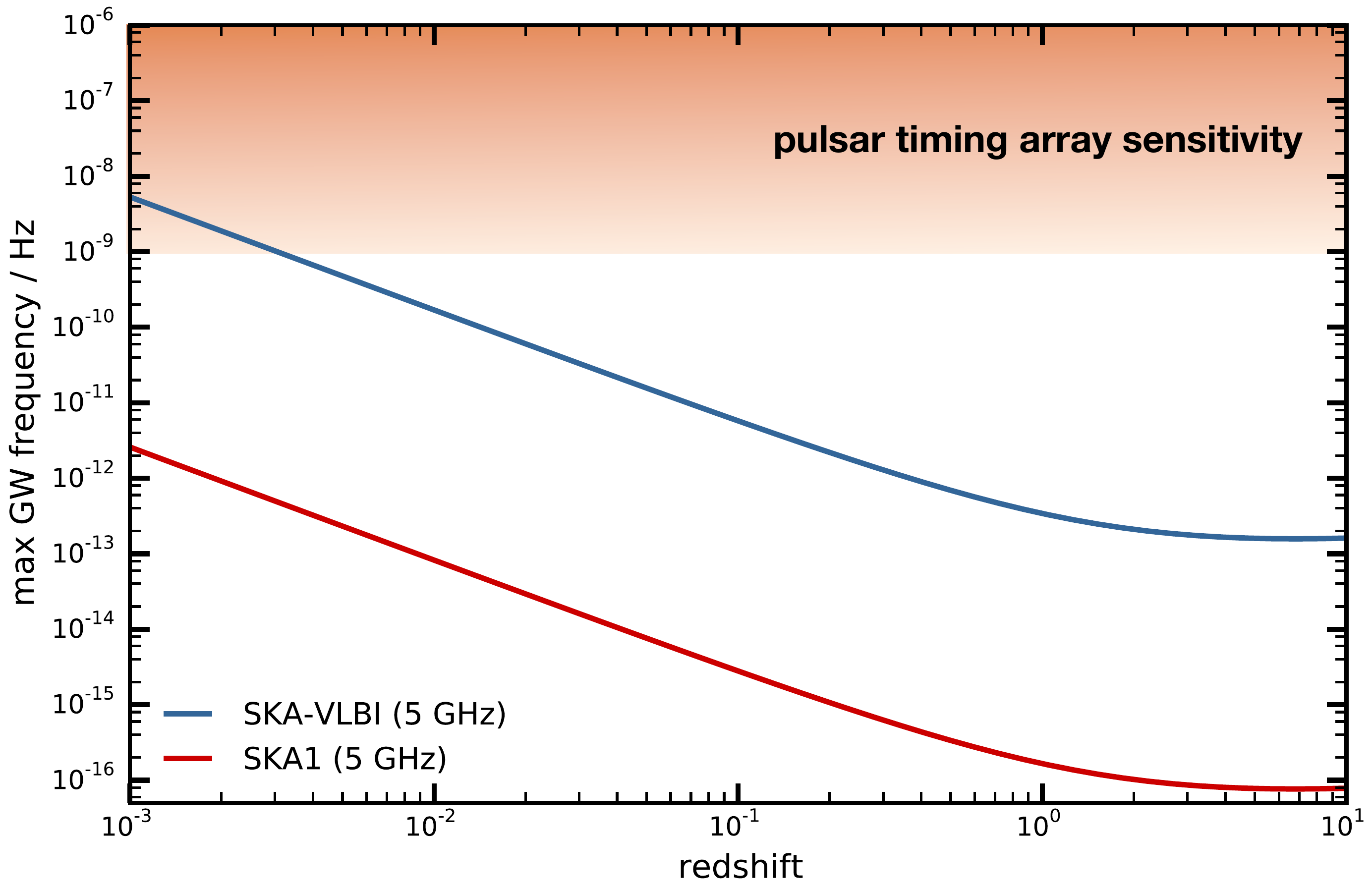} 
\end{center}
\caption{Maximum gravitational wave frequency from a binary SMBH system that can be spatially-resolved by SKA1-MID/SUR and SKA-VLBI. We assume binaries with an angular separation equal to 2 PSF FWHM, masses of 10$^9$~M$_\odot$ and circular orbits. This plot demonstrates that even high sensitivity VLBI is unlikely to discover sources in the frequency band detectable by pulsar timing arrays, however, as discussed in Sec.~2.2, this may not be necessary to find so-called gravitational wave hotspots.}
\label{fig:gwfreq}
\end{figure}

Despite this increased discovery efficiency that would be enabled by the
SKA, Fig.~\ref{fig:gwfreq} suggests that direct imaging is unlikely to
yield a large number of candidates for pointed gravitational wave
experiments in the pulsar timing array band. However, as we discuss in the
following section, the large-scale radio-jet morphology may provide an
alternative method to discover close-pair binaries not resolvable at VLBI
resolution. We note that the SKA is likely to be a 50+ year project and so
the lower frequency limit in the pulsar timing band will increase in time.
Halving the lower frequency limit results in roughly a factor of 8 larger
cosmological volume `surveyed', so a future generation of astronomers may
re-plot Fig.~\ref{fig:gwfreq} with an updated lower frequency limit and
come to a different conclusion.

\subsection{Radio-jet morphology as a proxy}\label{sec:jets}

\begin{figure}
\begin{center}
\includegraphics[width=0.75\textwidth]{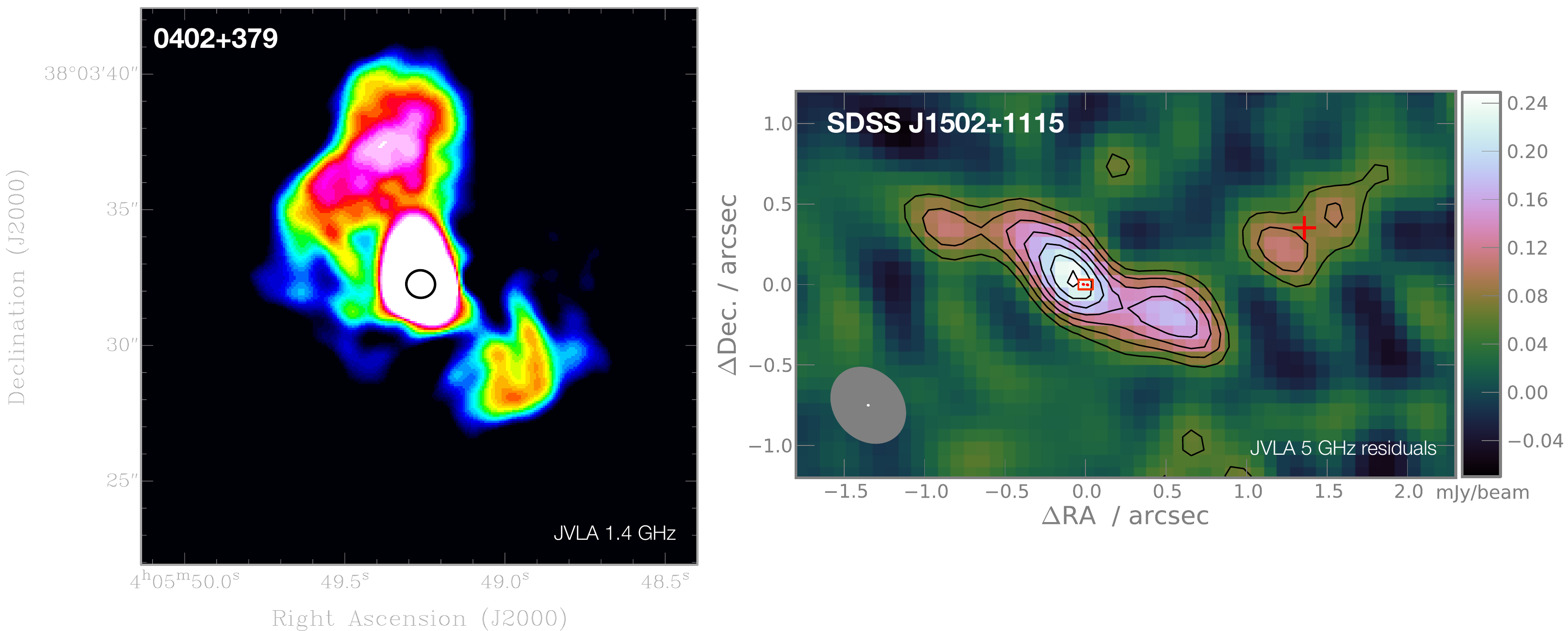} 
\end{center}
\caption{The above figures show the low surface brightness emission associated with the two lowest separation binary SMBH candidates (0402+379 and J1502+1115). This is to illustrate how arcsec-scale radio-jet morphology may reveal the presence of low separation binaries. {\bf Left panel: } JVLA 1.4~GHz map of 0402+379, the lowest separation ($\sim 7$~parsec) binary SMBH system known. The colour scale is clipped between 2-20~mJy\,beam$^{-1}$ (1$\sigma$ noise = 0.21 mJy\,beam$^{-1}$) to highlight the complex, potentially helical-like large-scale radio-jet emission centered on the VLBI cores which are within the 1.4 GHz PSF FWHM (shown by black ellipse in centre). VLBI observations resolve out all emission but the two radio cores and the innermost, few-parsec scale jets. Intermediate baselines ($\sim$200~km) are required to map the substantial 1.4~GHz jet emission at higher resolution. Figure from Deane et al.~(in prep.). {\bf Right panel: }Point-source subtracted JVLA 5~GHz residuals revealing rotationally symmetric "S"-shaped radio emission centred on VLBI components (within the red rectangle). There also appear to be small-scale jets centred on the third SMBH in the system to the west (J1502P, red cross). Over-plotted are the JVLA 5~GHz contours (black) starting at $\pm$60~$\mu$Jy\,beam$^{-1}$ ($\pm4\sigma$) and increasing/decreasing in steps of $\pm2\sigma$. The grey ellipse shows the JVLA 5~GHz PSF, while the white ellipse shows the 5~GHz EVN PSF. Modified figure from Fig.~1c in Deane et al.~(2014). } 
\label{fig:corkscrews}
\end{figure}

Even though SKA1-MID/SUR is only likely to detect dual AGN separations of $\lesssim$~1~kpc {\it for all redshifts}, it may be able to select excellent close-pair ($<<$~1~kpc) binary candidates via the large-scale radio-jet morphology, particularly given the unprecedented low surface brightness sensitivity. A number of studies have investigated the binary SMBH imprint on radio-jet morphology (e.g. Begelman et al. 1980, Kaastra \& Roos~1992). Deane et al.~(2014) show evidence of a $\sim$140~parsec binary SMBH in SDSS\,J1502+1115 using VLBI observations, as well as rotationally-symmetric "S"-shaped radio emission centered on the VLBI flat-spectrum cores. This suggests that the close-pair ($<<$1~kpc) binary SMBH could in principle be discovered using the rotationally-symmetric modulation imprinted onto the radio jets. The radio jets for the two lowest separation, spatially-resolved binary candidates (0402+379 and SDSS\,1502+1115) are shown in Fig.~\ref{fig:corkscrews}, indicating the binary orbits may have a significant influence on their morphology. As discussed in the introduction, hydrodynamical simulations suggest that the separations of $\sim$10$^8$~M$_\odot$ binary SMBHs can decrease from $\sim$100 parsec down to the gravitational radiation dominated regime ($<$1~parsec) on a timescale of a few Myr, particularly if sufficient gas is present (e.g. van Wassenhove et al.~2012). This is comparable to the typical lifetime of radio jets ($\sim$10~Myr), suggesting that these helical relics may provide arcsecond-scale signposts of close-pair binaries, some of which may not even be resolvable using VLBI. If that is the case, this may be a way to find gravitational wave `hotspots' where a nearby system with sufficiently high mass black holes may stand out above the stochastic gravitational wave background (e.g. Sesana 2014, Simon et al.~2014). Such targets would be very useful in understanding electromagnetic counterparts of gravitational wave sources. Therefore, radio-jet morphology may be a promising tool in binary SMBHs searches of the future, particularly given the expected SKA imaging fidelity and sensitivity at VLBI-scales. The above applies to `static' jets (negligible variation over the timescale of years), however, monitoring variations of parsec-scale jets (e.g. Agudo et al.~2012) is an additional probe, particularly given the cadence that jet position angles may be measured at high angular resolution. These two radio jet morphology techniques may identify promising targets for millimetre VLBI observations with $<$40~$\mu$-arcsecond angular resolution (e.g. Doeleman et al., 2008). This would in principle enable the direct imaging of a $z \lesssim 0.1$ binary SMBH that generates gravitational waves in the pulsar timing array band.

\subsection{Binary SMBH insights from pulsar timing arrays} 

As described in Janssen et al.~(2014, these proceedings), detection of the stochastic gravitational wave background is likely by the end of the decade. Pulsar timing arrays are sensitive to low-frequency (nHz-$\mu$Hz) gravitational waves, limited roughly by the length of time astronomers have been accurately timing millisecond pulsars at the low end ($\sim$1-3 decades); and by the required time to achieve sufficient timing residuals signal-to-noise (of the full network of pulsars) at the high end ($\sim 2$ weeks). This frequency range is thought to be dominated by binary SMBHs and therefore, such a detection will prove the existence of a population of sub-pc separation systems of high mass ($\gtrsim 10^7$~M$_\odot$). While pulsar timing arrays, radio-jet morphology and direct imaging of flat-spectrum sources are very different probes of binary SMBHs, these approaches will have to form a consistent picture. Each should in principle contribute to an overall understanding of the binary in-spiral evolution. 

For binary separations and masses that correspond to $\sim$nHz frequency, environment coupling is thought to be particularly important in determining the in-spiral rate and orbital eccentricity -- both of which have significant effects on the spectral shape of the stochastic gravitational wave background (as discussed in Sec.~\ref{sec:gw}). Therefore, the spectral shape of the nHz detected stochastic gravitational wave background must be reconciled with the merger rate and constraints of radio-detected AGN fraction at sub-kpc scales. Measurements of stellar core deficits with the JWST and E-ELT, as well as ALMA-derived molecular gas kinematics will be critical ancillary data products for SKA-discovered binaries. Variability and transient events may indicate mergers and a synergy between LSST and SKA is also expected in this particular pursuit (see Fender et al.~2014, these proceedings, and references therein).

What sets the SKA apart from the major multi-wavelength facilities of the future is the dynamic range of binary black hole separation that it will probe. This will be several ($\gtrsim$6) orders of magnitude when considering the virial radius of a large elliptical galaxy ($\sim$10~kpc) down to the sub-pc separations of binaries in the gravitational radiation dominated zone.

\section{Predictions for the SKA}

As Fig.~\ref{fig:status} illustrates, very few of the expected population of low separation binary SMBHs have been identified. Extrapolations to the number of systems the SKA will detect are therefore highly unconstrained. However, theoretical expectations suggest that 0.1-1 percent of galactic halos may host binary SMBHs, at least at intermediate to high redshift (e.g. Volonteri et al.~2003, Kulkarni \& Loeb 2012). Given the enhanced AGN triggering at low separation, as well as the high number density of radio AGN to be detected with the SKA continuum surveys ($\sim 10^4$ per square degree at 1.4 GHz; $> 5 \sigma$), we expect several orders of magnitude increase in the number of known binary SMBHs. We make no attempt to extrapolate from small number statistics, since the overwhelming message is that both phases of the SKA will open up a massive discovery space and will allow robust statistics of sub-kpc binary/dual AGN to be performed for the first time. This will enable constraints on the binary SMBH coalescence rate and correlations with their host galaxies via other multi-wavelength tracers to be explored. This revolutionary view of binary SMBHs in the Universe is critically dependent on high sensitivity on long baselines ($>200$~km) and VLBI capability for high angular resolution followup of the $\gtrsim 50 \, \mu$Jy sources. Here SKA-VLBI refers to combining existing VLBI networks with the SKA in both phases.

\subsection{SKA1 impact}

Assuming the above indicative estimate on the sub-kpc AGN prevalence, we should in principle be able to detect a large fraction of those with radio counterparts with SKA1-MID/SUR sensitivity. This simple extrapolation therefore predicts that SKA1-MID/SUR will make of order a few to a few hundred dual AGN detections per square degree. This will be a unique sample not only in size, but also in selection since it will be unbiased by dust and gas obscuration. Furthermore, SKA1-MID/SUR will be able to probe these scales (1-100~kpc) at all redshifts (see Fig.~\ref{fig:status}) and so will become a dominant driver of this field. A 50 percent decrease in SKA1-MID/SUR sensitivity will only impact the science output through the rough factor of $\sqrt{2}$ decrease in the number of dual AGN detected, assuming a comparable baseline distribution.  

SKA1 is also expected to participate in VLBI sessions with existing networks (e.g. European and African VLBI Networks; Paragi et al.~2014, these proceedings). SKA1-MID/SUR will therefore already provide the opportunity to perform high sensitivity observations at milli-arcsecond angular resolution for a few weeks per year.

\subsection{SKA2 impact}

SKA2 will naturally lead to a significant increase in the sample of known parsec-scale binary AGN, through the dramatic increase in sensitivity on longer baselines ($>$~200~km) for the majority of the available telescope observing time. SKA2-VLBI will be able to resolve binaries with separations of $\sim20$~pc at all redshifts and map associated jets with high fidelity. The  wide-area ($\gtrsim$10~deg$^2$) mas-resolution surveys will open a completely new parameter space and discover a large number of systems of comparable and lower separation that the current record holder: 0402+379, the 7~pc separation binary at $z\sim0.06$ (Rodriguez et al.~2006). This increased sample size at low separation will be key in bridging the black hole coalescence rate and environment coupling with the results from pulsar timing arrays. The latter requires the stochastic gravitational wave background nHz spectral shape to be measured from pulsar timing arrays, which may be possible by the time SKA2 is complete. In addition, these surveys will provide a large sample of high precision measurements of AGN positions, enabling a statistical study of the offset AGN predicted by gravitational recoil or the presence of binary SMBHs (e.g. Orosz \& Frey 2013, Paragi et al., these proceedings).

This suggests that SKA2 (particularly when combined with existing VLBI arrays) will completely revolutionize this field once again (following SKA1-MID/SUR) in the low separation parameter space and bridge pulsar timing array results with what is gleaned from mas-scale continuum surveys. This naturally leads to the question: {\sl will SKA2-VLBI resolve SMBH binaries that can be detected by pulsar timing arrays?} In Fig.~\ref{fig:gwfreq} we have shown the maximum gravitational wave frequency of a binary that has a 10 milliarcsecond separation, mass of $2 \times 10^9$~M$_\odot$, and circular orbit. This demonstrates that even SKA2-VLBI is unlikely to discover binaries in the pulsar timing array sensitivity frequency band at 1-2~GHz. This places greater emphasis on higher frequency surveys and searching for low separation (sub-pc) binaries using the large-scale radio-jet morphology. We note the exciting possibility that the SKA may discover such systems purely through pulsar timing array experiments, however, the associated positional uncertainties ($\sigma_{\rm pos} >> 10$~deg$^2$) may prohibit electromagnetic counterpart determination.

\section{Summary}

This contribution discusses how the SKA will constrain the black hole merger rate at low separations ($<<$1~kpc). The preliminary evidence that radio-jet triggering is enhanced in dual and triple SMBH systems; negligible dust and gas attenuation at GHz frequencies; and unmatched sensitivity and angular resolution suggests that the SKA will play a leading role in opening up the low separation parameter space. Identification of galaxies with binary SMBH systems by the SKA will have significant multi-wavelength synergy with current/future facilities such as ALMA, E-ELT, JWST and LSST, all of which will contribute to understanding the astrophysical impact thereof. Also of great importance will be the ability of high brightness temperature sensitivity observations to probe relic emission from precessing jets and hence provide arcsec-scale proxies of close-pair binary SMBH systems that cannot not be spatially resolved by VLBI. If sufficiently massive binaries are found, these may be detectable by directed gravitational wave experiments with pulsar timing arrays. The range of science possible with SMBHs is extensive and the SKA will play a leading role in their discovery, characterization and followup. These different observing modes are therefore expected to usher in the exciting era of multi-messenger astrophysics with a single facility.

\end{document}